\documentclass[aps,prx,twocolumn,superscriptaddress,noshowpacs,floatfix]{revtex4-2}
\usepackage{graphicx}
\usepackage{xcolor}
\usepackage[normalem]{ulem}
\usepackage{float}
\usepackage{bm}
\usepackage{braket}
\usepackage{placeins}
\usepackage{amsmath}
\usepackage{physics}
\usepackage{lipsum}
\usepackage{verbatim}
\usepackage{float}
\usepackage{ulem}
\usepackage{soul}

\newcommand{\red}[1]
{\textcolor{black}{#1}}
\newcommand{\af}[1]
{\textcolor{black}{#1}}

\newcommand{\mycomment}[1]{}

\begin{document}

\title{Possibility of ferro-octupolar order in Ba$_2$CaOsO$_6$ assessed by X-ray magnetic dichroism measurements}

\author{G.~Shibata}
\affiliation{Materials Sciences Research Center, Japan Atomic Energy Agency, Sayo, Hyogo 679-5148, Japan}

\author{N. Kawamura}
\affiliation{Japan Synchrotron Radiation Research Institute, Sayo, Hyogo 679-5198, Japan}

\author{J.~Okamoto}
\affiliation{National Synchrotron Radiation Research Center, Hsinchu 300092, Taiwan}

\author{A.~Tanaka}
\affiliation{Department of Quantum Matter, Hiroshima
University, Hiroshima 739-8530, Japan}

\author{H.~Hayashi}
\altaffiliation [present address: ] {\emph{Institute for Solid State Physics, The University of Tokyo, 5-1-5 Kashiwanoha, Kashiwa, Chiba 277-8581, Japan}}
\affiliation{Research Center for Materials Nanoarchitectonics (MANA),National Institute for Materials Science, Tsukuba, Ibaraki 305-0044,Japan}
\affiliation{Graduate School of Chemical Sciences and Engineering, Hokkaido University, Sapporo, Hokkaido 060-0810, Japan}

\author{K.~Yamaura}
\affiliation{Research Center for Materials Nanoarchitectonics (MANA),National Institute for Materials Science, Tsukuba, Ibaraki 305-0044,Japan}
\affiliation{Graduate School of Chemical Sciences and Engineering, Hokkaido University, Sapporo, Hokkaido 060-0810, Japan}

\author{H.~Y.~Huang}\affiliation{National Synchrotron Radiation Research Center, Hsinchu 300092, Taiwan}

\author{A.~Singh}\altaffiliation [present address: ]{\emph{Department of Physics and Astrophysics, University of Delhi, Delhi 110007, India}}\affiliation{National Synchrotron Radiation Research Center, Hsinchu 300092, Taiwan}

\author{C.~T.~Chen}\affiliation{National Synchrotron Radiation Research Center, Hsinchu 300092, Taiwan}

\author{D.~J.~Huang}
\affiliation{National Synchrotron Radiation Research Center, Hsinchu 300092, Taiwan}
\affiliation{Department of Physics, National Tsing Hua University, Hsinchu 300044, Taiwan}
\affiliation{Department of Electrophysics, National Yang Ming Chiao Tung University, Hsinchu 300093, Taiwan}

\author{S.~V.~Streltsov}
\affiliation{Institute of Metal Physics, 620041 Ekaterinburg GSP-170, Russia}

\author{A.~Fujimori}
\affiliation{Materials Sciences Research Center, Japan Atomic Energy Agency, Sayo, Hyogo 679-5148, Japan}
\affiliation{National Synchrotron Radiation Research Center, Hsinchu 300092, Taiwan}
\affiliation{Department of Physics, National Tsing Hua University, Hsinchu 300044, Taiwan}
\affiliation{Department of Physics, University of Tokyo, Bunkyo-Ku, Tokyo 113-0033, Japan}

\begin{abstract}

Localized $5d^2$ electrons in a cubic crystal field possess multipoles such as electric quadrupoles and magnetic octupoles. We studied the cubic double perovskite Ba$_2$CaOsO$_6$ containing the Os$^{6+}$ ($5d^2$) ions, which exhibits a phase transition to a `hidden order' below $T^* \sim$ 50 K, by X-ray absorption spectroscopy (XAS) and X-ray magnetic circular dichroism (XMCD) at the Os $L_{2,3}$ edge. 
The cubic ligand-field splitting between the $t_{2g}$ and $e_g$ levels of Os $5d$ was deduced by XAS to be $\sim$4 eV. 
Ligand-field (LF) multiplet calculation under fictitious strong magnetic fields indicated that the exchange interaction between nearest-neighbor octupoles should be as strong as $\sim$1.5 meV if a ferro-octupolar order is stabilized in the `hidden-ordered' state, consistent with the exchange interaction of $\sim$1 meV 
previously predicted theoretically using model and density functional theory calculations.
The temperature dependence of the XMCD spectra was consistent with a $\sim$18 meV residual cubic splitting of the lowest $J_{\rm eff} =$ 2 multiplet state into the non-Kramers $E_g$ doublet ground state and the $T_{2g}$ triplet excited state.

\end{abstract}
\date{\today}
\maketitle

\section{Introduction}

\begin{figure}[tb]
\centering
\includegraphics[width=1.0\columnwidth]{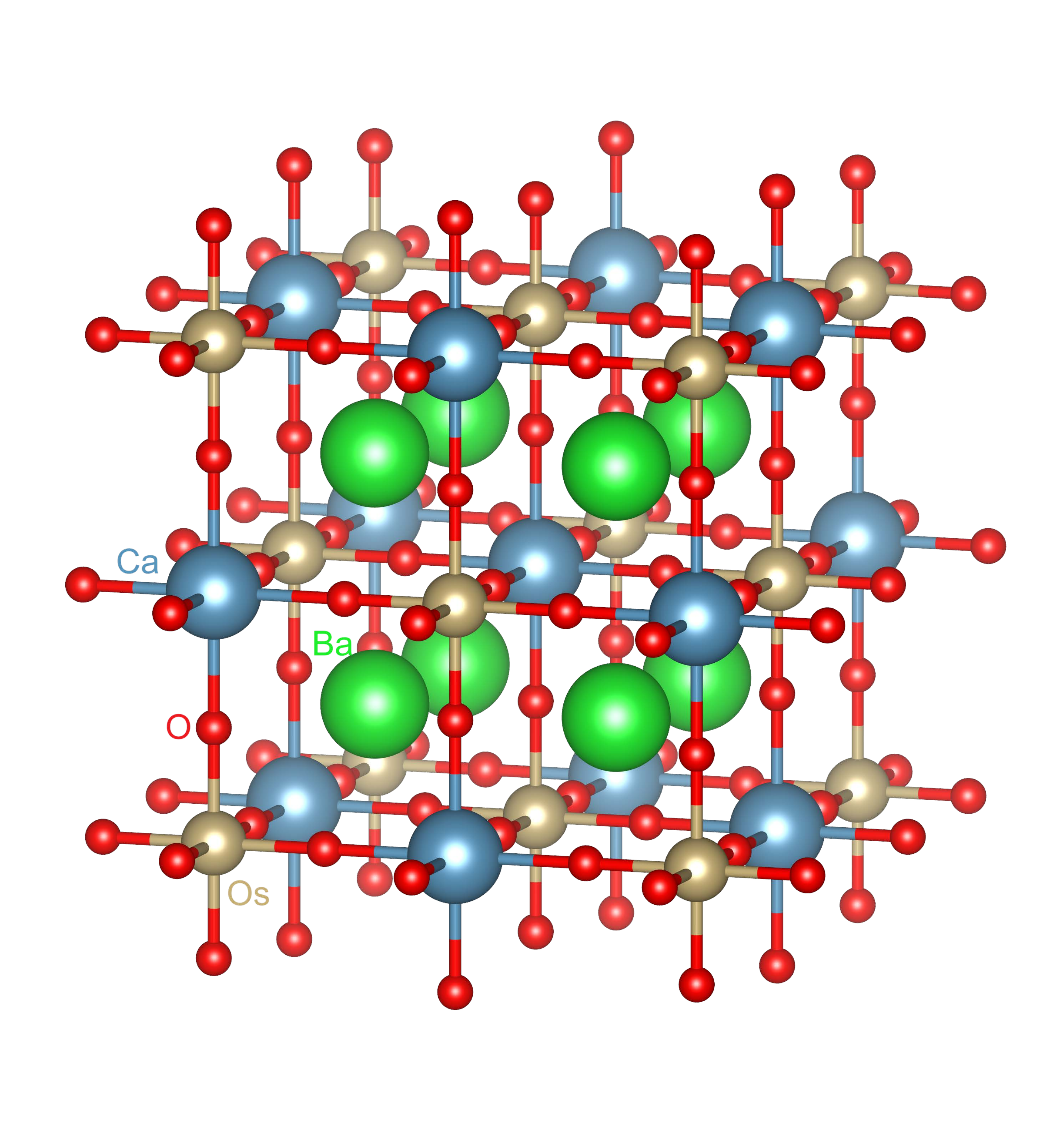}
\caption{\af{Double-perovskite} \red{crystal structure of Ba$_2$CaOsO$_6$ drawn using the VESTA software \cite{vesta}. The Os-Os and Os-O distances are estimated to be 5.9012 \AA \ and 1.9109 \AA, respectively \cite{maharaj}.}}
\label{crysstruct}
\end{figure}

Correlated electronic states in strongly spin-orbit coupled systems have attracted strong interest in recent years~\cite{Khomskii2021,KHOMSKII-enc}. Particular attention has been attracted by the $J_{\rm eff} =$1/2 Mott insulators Ir$^{4+}$ ($5d^5$) oxides~\cite{bjkim-sjoh,bjkim-arima} and their doped compounds~\cite{ykkim-bjkim} as well as the Kitaev quantum-spin-liquid candidates Ru$^{3+}$ ($4d^5$) honeycomb compounds~\cite{suzuki-keimer}. Recently, it was theoretically predicted that a magnetic octupolar order can occur in localized $5d^2$ electrons is a cubic crystal field~\cite{chen-balents,svodova,paramekanti,khaliullin}. In general, a $d^2$ ion coordinated by ligand atoms in the cubic ($O_h$-symmetry) environment is expected to undergo a Jahn-Teller distortion, but many cubic crystals with $5d^2$ ions remain undistorted owing to the strong spin-orbit coupling (SOC) of the $5d$ electrons~\cite{streltsov}. This drew the attention of many researchers to $B$-site-ordered double-perovskite oxides containing $5d^2$ ions (Os$^{6+}$, Re$^{5+}$) such as Ba$_2B$OsO$_6$ and Ba$_2B'$ReO$_6$, where \textit{B} is an alkali earth and \textit{B$'$} is a rare earth. 
\red{Figure \ref{crysstruct} shows the crystal structure of the $B$-site-ordered double-perovskite oxides Ba$_2$CaOsO$_6$; the OsO$_6$ octahedra exhibit $O_h$ symmetry and are isolated from one another, which enables study the intrinsic electronic structures of $5d^2$ systems.}

The magnetic susceptibility of Ba$_2$CaOsO$_6$ containing Os$^{6+}$ ions exhibits a cusp-like anomaly at $T^\ast \sim$ 50 K, but neutron diffraction has shown no magnetic Bragg peaks below $T^\ast$ while muon-spin rotation ($\mu$-SR) has revealed a small local magnetic moment of $\sim$ 0.2 $\mu_{\rm B}$ ~\cite{thompson} and a staggered magnetic moment of $\sim$ 0.05 $\mu_{\rm B}$ per Os ion~\cite{cong}. According to X-ray diffraction, the crystal remains cubic down to the lowest temperatures, precluding an electric quadrupolar order that should lead to a Jahn-Teller distortion~\cite{hirai}. Therefore, the origin of the `hidden order' below $T^\ast$ is consistent with an ordering of the magnetic octuploes. Theoretically, exchange coupling between neighboring Os ions favors ferro-octupolar order in the fcc sublattice of the double perovskite ~\cite{paramekanti}. 

\begin{figure}[tbp]
\centering
\includegraphics[width=1.0\columnwidth]{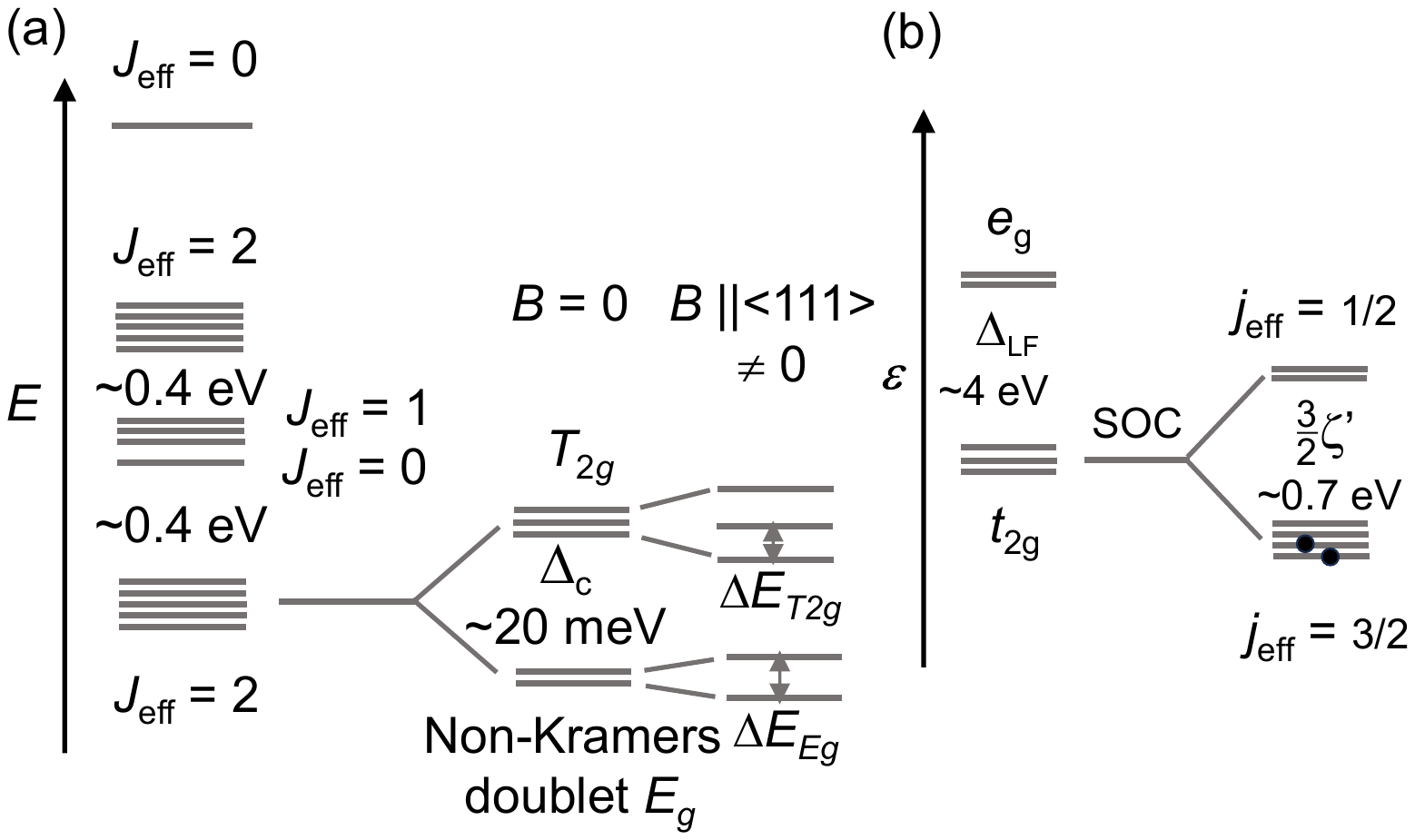}
\caption{Energy levels of the Os$^{6+}$ ($5d^2$) ion in the cubic ($O_h$) crystal field. 
(a) The $t_{2g}^2$ part of the $5d^2$ multiplet levels. The $t_{2g}e_g$ part is located at higher energies separated by the $t_{2g}$-$e_g$ splitting of $\Delta_{\rm LF}\sim 4$ eV, see panel (b). The lowest multiplet state $J_{\rm eff}= 2$ is split by a residual cubic splitting, $\Delta_{\rm c}$, into the non-Kramers $E_g$ doublet ground state and the $T_{2g}$ triplet excited states. Under a finite magnetic field $B\parallel \langle 111\rangle$, the non-Kramers doublet is split into the two magnetic octupolar eigenstates $|\psi_{g,\pm}\rangle$ (defined in the text) separated by $\Delta E_g$ ($\propto B^3$) and the triplet is split by the Zeeman energy $\Delta E_{T_{2g}}$  ($\propto B$).
(b) One-electron energy levels of the Os $5d$ orbitals. $\Delta_{\rm LF}$ is the  $t_{2g}$-$e_g$ ligand-field splitting. Spin-orbit coupling (SOC) splits the $t_{2g}$ level further into the $j_{\rm eff} = \frac{1}{2}$ and $j_{\rm eff} = \frac{3}{2}$ levels separated by $\frac{3}{2}\zeta'$, 
and the $j_{\rm eff} = \frac{3}{2}$ level is occupied by two electrons.}
\label{ediagram}
\end{figure}

The formation of the magnetic octupole from the $5d^2$ configuration in a cubic crystal field is illustrated in Fig.~\ref{ediagram}(a) based on the two-electron energy diagram that has been obtained by recent resonant inelastic X-ray scattering (RIXS) studies~\cite{okamoto,Frontini2}. 
Under the $O_h$ symmetry, the lowest state of the $t_{2g}^2$ multiplet, which has the total effective angular momentum of $J_{\rm eff} = 2$, should split into the non-Kramers $E_g$ doublet ground state and the triply degenerate $T_{2g}$ excited state, separated by a residual cubic splitting $\Delta_{\rm c}$~\cite{maharaj,Voleti2020}. 
Depending on the nature of additional perturbation on the Os$^{6+}$ ion, the non-Kramers doublet may split into either eigenstates of the electric quadrupole operators such as 
$|\psi_{g,\uparrow}\rangle\equiv |J^z_{\rm eff}=0\rangle$ and
$|\psi_{g,\downarrow}\rangle\equiv \frac{1}{\sqrt{2}}\big(|J^z_{\rm eff}=2\rangle+|J^z_{\rm eff}=-2\rangle\big)$, or eigenstates of the magnetic octupole operator $T\propto \overline{J^xJ^yJ^z}$ (the overline denotes symmetrization), such as  $|\psi_{g,\pm}\rangle\equiv\frac{1}{\sqrt{2}}\big(|\psi_{g,\uparrow}\rangle \pm i|\psi_{g,\downarrow}\rangle\big)$.
If all the Os ions are in one of the two eigenstates $|\psi_{g,\pm}\rangle$ of the $T$ operator, the ferro-octupole-ordered state is realized. In order to investigate the ground state and low-energy excited states of the Os$^{6+}$ ion, which are more directly related to the multipole orders, linear or circular dichroism of X-ray absorption spectroscopy (XAS) and its temperature dependence are expected to provide us with valuable information.

In this work, we measured XAS and X-ray magnetic circular dichroism (XMCD) at the Os $L_{2,3}$ edges of Ba$_2$CaOsO$_6$ and analyzed the spectra using ligand-field (LF) multiplet theory. In particular, we addressed the question of why the non-Kramers $E_g$ doublet prefers the magnetic octupole to the electric quadrupole as the ground state on the basis of the experimental and calculated low-energy states of the Os$^{6+}$ ion under magnetic fields. In particular, multiplet calculation under fictitious strong magnetic fields suggested that the exchange interaction between nearest-neighbor Os ions should be as strong as $\sim$1.5 meV if the hidden order is caused by a ferro-octupolar order. This is consistent with the recent density-functional theory (DFT) calculation that predicted the exchange interaction of $\sim$1 meV ~\cite{voleti-tanusri}.

\section{results}

\begin{figure}[tbp]
\centering
\includegraphics[width=1.0\columnwidth]{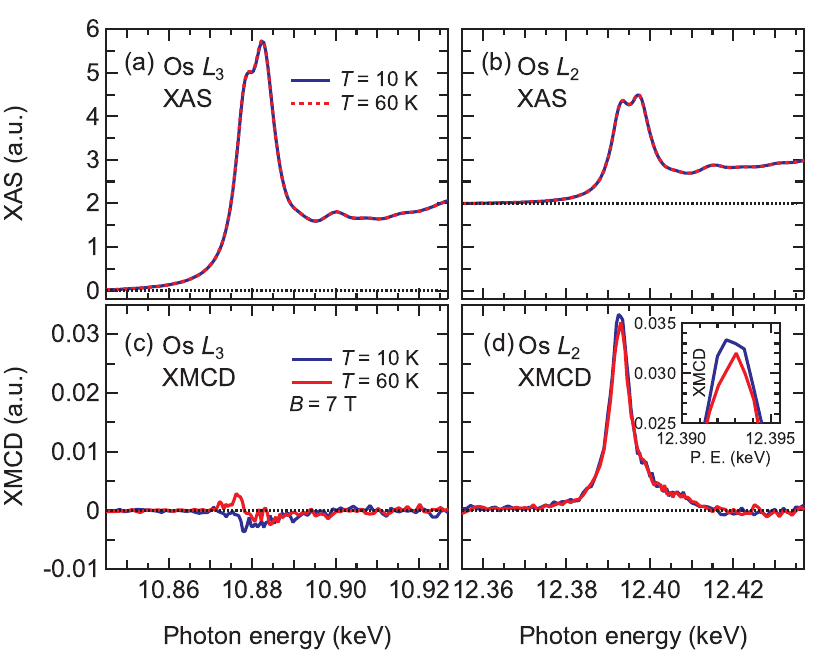}
\caption{X-ray absorption spectroscopy (XAS) and x-ray magnetic circular dichroism (XMCD) spectra of Ba$_2$CaOsO$_6$ at the Os $L_{2,3}$ edges measured at $T=10\ \text{K}$ and $60\ \text{K}$ under the magnetic fields of $\pm$7 T  before background subtraction. (a), (b) XAS spectra at the Os $L_3$ and $L_2$ edges, respectively. (c), (d) XMCD spectra at the Os $L_3$ and $L_2$ edges, respectively. 
\red{Inset shows the expanded XMCD spectra at the Os $L_2$ edge. }
}
\label{xasxmcd}
\end{figure}

Figure~\ref{xasxmcd} shows XAS and XMCD spectra taken at $T=10\ \text{K}$ and $60\ \text{K}$ under the magnetic fields of $B =\pm$ 7 T. 
 (For experimental details, see Appendix~\ref{xmcd}.)
 The XAS spectra show a double-peak structure both at the $L_3$ and $L_2$ edges [Figs.~\ref{xasxmcd}(a) and (b), respectively], reflecting the $t_{2g}$-$e_g$ ligand-field splitting of $\Delta_{\rm LF}\sim$ 4 eV of the Os $5d$ level [defined in Fig.~\ref{ediagram}(b)]. 
This $\Delta_{\rm LF}$ value is nearly identical to the value deduced from the O $K$-edge XAS measurement~\cite{okamoto}. 
Only the XAS peaks arising from transition to the $t_{2g}$ states show finite XMCD signals. The XMCD intensity at the Os $L_{2}$ edge  [Fig.~\ref{xasxmcd}(d)] is high ($\sim 1\%$ of XAS) while that at the Os $L_{3}$ edge [Fig.~\ref{xasxmcd}(c)] is extremely low ($\lesssim 0.1\%$). 
Such highly $L_3$-$L_2$ asymmetric XMCD intensities have also been observed in some ferrimagnetic double-perovskite oxides containing Os~\cite{MorrowXMCD,VeigaXMCD}. 
The XMCD sum rules yield the induced orbital and `effective' spin magnetic moments (see Appendix~\ref{xmcd}) of 
$M_\text{orb}=-0.016 \af{\pm 0.002} \mu_\text{B}/{\rm Os}$ and $M_\text{spin}^\text{eff}=0.058 \af{\pm 0.007} \mu_\text{B}/{\rm Os}$ 
at $T=10\ \text{K}$, and $M_\text{spin}^\text{eff}$ decreases by \af{$\sim$}0.004 $\mu_{\rm B} $ at $T=60\ \text{K}$. 
The relatively large unquenched orbital magnetic moment suggests that the strong SOC of Os $5d$ orbitals indeed plays an important role in the electronic structure of Ba$_2$CaOsO$_6$. 

\begin{figure}[tbp]
\centering
\includegraphics[width=1.0\columnwidth]{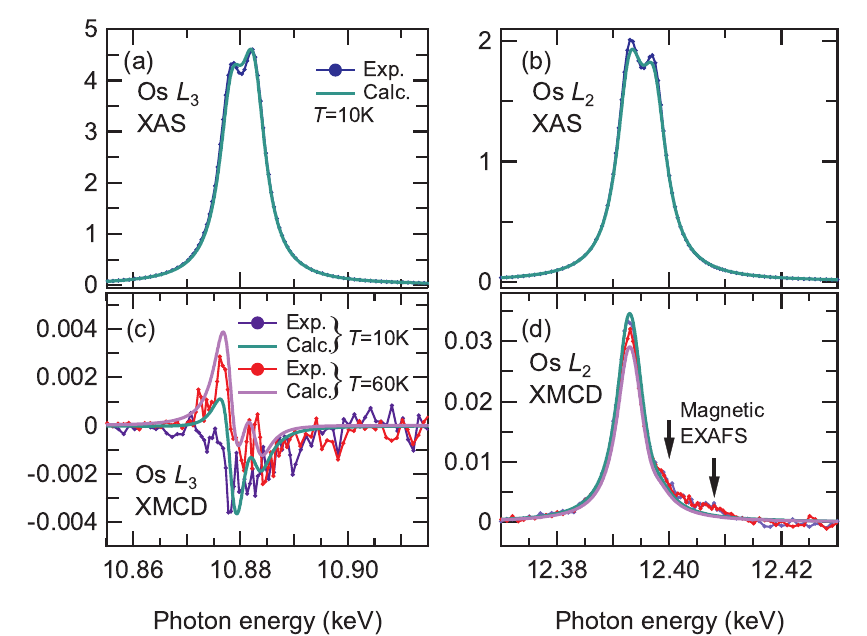}
\caption{XAS and XMCD spectra of Ba$_2$CaOsO$_6$ at the Os $L_{2,3}$ edges after background subtraction compared with ligand-field (LF) multiplet calculation.
(a), (b) Experimental (blue curves with dots) and calculated (green curves) XAS spectra. The white-line background and the extended X-ray absorption fine structure (EXAFS) oscillations have been subtracted (see Appendix~\ref{xmcd}). (c), (d) Experimental (purple and red curves with dots) and calculated (green and lilac curves) XMCD spectra. The structures around the energies of 12.40 and 12.41 keV in the experimental XMCD spectra are magnetic EXAFS oscillations\af{, the period of which coincides with that of EXAFS in Ba$_2$Na$_{1-x}$Ca$_x$OsO$_6$~\cite{Kesavan}.} The molecular field in the multiplet calculation was adjusted to $B =$ 12 $\pm 2\ \text{T}$ so that the measured XMCD intensity at the $L_2$ edge was reproduced.}
\label{tdep}
\end{figure}

Figure~\ref{tdep} shows comparison of the observed Os $L_{2,3}$-edge XAS spectra after background subtraction and the XMCD spectra with LF multiplet calculation. 
 (Details of the background subtraction are described in Appendix~\ref{xmcd}.)
We used the ligand-field splitting of $\Delta_{\rm LF}= $ 4.3$\ \text{eV}$ and the SOC parameter of $\zeta = $ 0.33$\ \text{eV}$ to achieve the best fit. 
Here, $\zeta$ $(\equiv -\zeta')$ is defined by the SOC energy $\zeta({\bf l}\cdot {\bf s})=\zeta'({\bf l}_{\rm eff}\cdot {\bf s})$, where ${\bf l}$ is the orbital angular momentum of an Os $5d$ electron and ${\bf l}_{\rm eff} \equiv -{\bf l}$ is the effective angular momentum of a $t_{2g}$ electron~\cite{kanamori}. 
To reproduce the measured XMCD intensities, we had to assume an effective molecular field of $B=$ 12$\pm$2 T on top of the external magnetic field of 7 T in the calculation [Figs.~\ref{tdep}(c) and (d)].
The experimental value of $|M_\text{orb}/M_\text{spin}^\text{eff}|=0.28 \af{\pm 0.04}$ deduced using the XMCD sum rules under the external magnetic field of $B=$ 7 T was reproduced by the theoretical value of $|M_\text{orb}/M_\text{spin}^\text{eff}|=$ 0.30\af{$\pm$0.04} value calculated for 10 K under $B=$ 7 T plus the molecular field of 12$\pm$ 2 T.
This supports our initial assumption that the Os $5d$ electrons in Ba$_2$CaOsO$_6$ are basically localized in spite of the strong hybridization with the O $2p$ orbitals. 
Using the multiplet calculation, we could separate $M_{\rm spin}^{\rm eff}$ into the spin magnetic moment $M_{\rm spin}$ and the magnetic dipole $M_T$:  $M_\text{spin}/M_\text{spin}^\text{eff} = 0.45 \af{\pm 0.05}$ and $M_T \equiv \frac{2}{7}(M_{\rm spin}^{\rm eff} - M_{\rm spin}) = 0.16 \af{\pm 0.02}$ (Appendix~\ref{xmcd}). 
As $M_\text{T}/M_\text{spin}^\text{eff}$ is a measure of anisotropic spin distribution on the Os ion~\cite{TXMCD_Stohr,TXMCD_Durr,shibata}, the large $M_\text{T}$ reflects the distortion of the $5d$ orbitals induced by spin polarization via strong SOC.

Under the external magnetic field of $B=$ 7 T employed for the XMCD measurements plus the effective molecular field of 12 T, the multiplet calculation predicts that the non-Kramers $E_g$ doublet is split by $\Delta E_{E_g}\sim$1 $\mu$eV, and the triplet excited states $T_{2g}$ should show a Zeeman splitting $\Delta E_{T_{2g}} =$ 0.74 meV, where $\Delta E_{T_{2g}}$ is defined in Fig.~\ref{ediagram}(a). 
The temperature dependence of the XMCD spectra \af{is compared with} the calculation in Figs.~\ref{tdep}(c) and (d). 
\af{Here, the calculated temperature dependence arises from the thermal excitation of the Os$^{6+}$ ion from the $E_g$ doublet ground state to the $T_{2g}$ triplet excited state across the residual cubic splitting $\Delta_{\rm c}$, which was calculated to be 18 meV using the above parameter set, and is consistent with the previous neutron scattering experiment~\cite{maharaj}.}

The XMCD spectra and their temperature dependence are consistent with the non-Kramers $E_g$ doublet ground state and the $T_{2g}$ triplet excited state separated by $\Delta_{\rm c}\sim 18$ meV.
Because the previous work show no evidence for Jahn-Teller distortion~\cite{okamoto}, the non-Kramers $E_g$ doublet ground state, which can host either electric quadrupole or magnetic octupole, most likely chooses the magnetic octupolar state~\cite{chen-balents,paramekanti,svodova,khaliullin}.
\af{Unfortunately, the Os $L_3$-edge XMCD spectra predicted by the multiplet calculation are very similar between the magnetic octupolar and electric quadrupolar $E_g$ states.}
Well below the transition temperature $T^\ast\sim$ 50 K, therefore, one may consider that the non-Kramers doublet is split into two octupolar states $|\psi_{g,\pm}\rangle$ separated by $\Delta E_{E_g}\sim k_{\text{B}}T^\ast\sim 4$ meV. Such a splitting would be caused by an effective molecular field due to interaction with neighboring Os ions.  
Our multiplet calculation showed that, magnetic fields along $\langle 111\rangle$ directions create purely octupolar states out of the non-Kramers doublet, which is consistent with the scenario that exchange interaction between the Os ions stabilizes the octupolar order. 
Because our multiplet calculation showed $\Delta E_{E_g}\sim 1$ $\mu$eV at $B=$ 7 T plus the effective molecular field of 12$\pm$2 T during the XMCD measurements and because $\Delta E_{E_g}\propto B^3$, $\Delta E_{E_g}\sim 4$ meV would be realized if the internal exchange field were as strong as $B \sim$ 300 T. 

\section{discussion}

\begin{figure}[tbp]
\centering
\includegraphics[width=0.9\columnwidth]{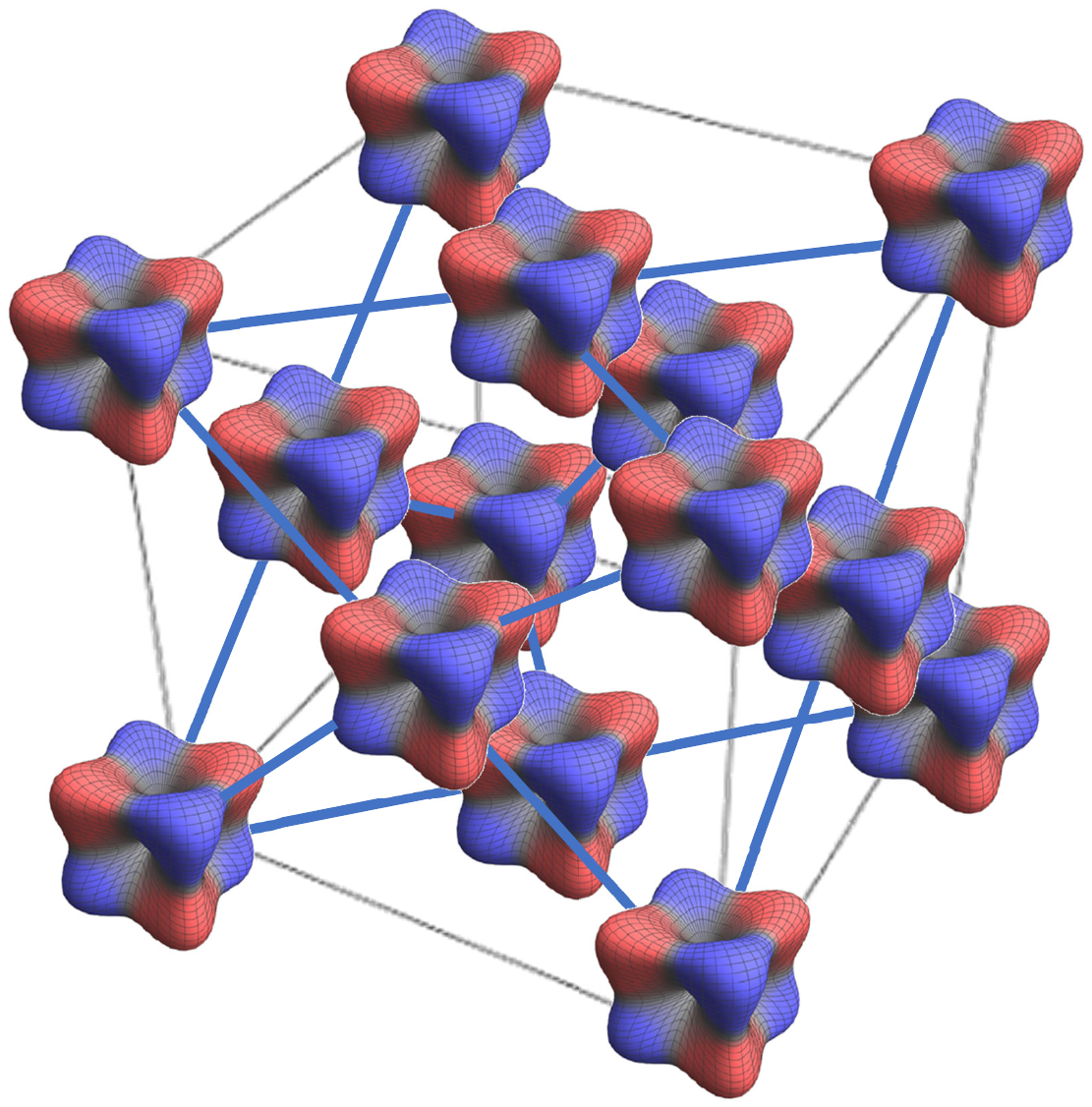}
\caption{Schematic drawing of the ferro-octupolar order of the Os$^{6+}$ ions in Ba$_2$CaOsO$_6$. Red and blue colors indicate the distribution of spin-up and spin-down electrons, respectively. Nearest-neighbor Os atoms are connected by blue lines.}
\label{octorder}
\end{figure}

\af{The magnetic octupole moment, i.e., the expectation value of the $T$ operator is calculated to be 1.26, which is nearly independent of $B$ as long as the sign of $B$ does not change.} If we assume that ferro-octupolar order is realized in Ba$_2A$OsO$_6$ as theoretically predicted~\cite{paramekanti}, exchange interaction $J$ between nearest-neighbor Os ions should be as strong as $B \sim$ 300 T divided by the number of the nearest-neighbor Os ions in the fcc lattice, 12, as indicated in Fig.~\ref{octorder}, that is, $J$ should be as large as $\sim$ 1.5 meV. 

\af{The above $J$} value is in fair agreement with \red {previous theoretical calculations. In particular, direct calculation of exchange tensors for different multipoles by combination of DFT and the dynamical mean-field theory (DMFT) shows that $J$ is expected to be $\sim$ 0.75 meV \cite{pourovskii}, but this result depends on chosen Hubbard $U$ parameter. On another hand, the Schrieffer-Wolff transformation applied to the two-site Hubbard-like model including the spin-orbit coupling with parameters estimated by DFT calculations gives $J\sim$ 1 meV~\cite{voleti-tanusri}.
Finally, it has to be mention that the same octupolar exchange field} may induce a small magnetic moment through $E_g$-$T_{2g}$ hybridization, as detected by zero-field $\mu$SR~\cite{thompson,cong}. 

More direct proof of the octupolar order may be given by impurity doping that breaks the local symmetry and induces detectable phenomena such as the formation of magnetic dipoles localized near the impurities~\cite{voleti-tanusri}.  
Large angle neutron scattering, as has been applied to the Ce pyrochlores,  may be used for the direct observation of magnetic octupoles ~\cite{silbille,bhardway}. \af{Recently, non-linear Hall effect at high frequencies has been proposed as a probe of magnetic ferro-octupolar order~\cite{Sorn,He-Law}.} The latter two methods would become applicable to Ba$_2$CaOsO$_6$ when single crystals become available.

\section{Conclusion}

Although XMCD is not a direct probe of magnetic octupoles by principle, the present temperature-dependent XMCD study at the Os $L_{2,3}$ edge combined with ligand-field multiplet calculation revealed the splitting of the lowest $J_{\rm eff}=2$ state of the $t_{2g}^2$ multiplet of the Os$^{6+}$ ion into the non-Kramers $E_g$ doublet ground state and the the $T_{2g}$ triplet excited states separated by the residual cubic splitting $\Delta_{\rm c}\sim$ 18 meV. From our ligand-field multiplet calculation of the XMCD spectra, we concluded that, if the exchange field on the Os$^{6+}$ ion were as strong as $\sim$300 T in the ferro-octupolar ordered state, the non-Kramers double ground state would be split by $\sim$4 meV and may give the transition temperature of $T^\ast\sim$ 50 K to the `hidden ordered' state.

\section*{Acknowledgements}
We are grateful to M. Haverkort, C. Franchini, L. Pourovskii, and A. Paramekanti for useful discussions. The experiment at BL39XU of SPring-8 was performed with the approval of the Japan Synchrotron Radiation Research Institute (JASRI) under Proposal No.\ 2022A1610.  
This work was partly supported by the National Science and Technology Council of Taiwan under Grant Nos.~103-2112-M-213-008-MY3, 108-2923-M-548-213-001, and 113-2112-M-007-033 and by the Japan Society for the Promotion of Science under Grant Nos. JP20K14416, JP22K03535, JP23K11709, and JP25K01657. S.V.S. was supported by the Ministry of Science and Higher Education of the Russian Federation. 
A.F. acknowledges the support of the Yushan Fellow Program and the Center for Quantum Science and Technology within the framework of the Higher Education Sprout Project under the Ministry of Education of Taiwan.



\section*{Data Availability}

All the data that support the findings of this article are available upon reasonable request.

\appendix

\section{X-ray magnetic circular dichroism \af{measurements} and sum rule \af{analysis}}\label{xmcd}

\red{High-quality polycrystalline samples of Ba$_2$CaOsO$_6$ were synthesized and characterized as described in Ref.~\cite{okamoto}. The obtained samples were gray sintered pellets with a diameter of $\sim 5\ \text{mm}$ and a thickness of $\sim 2\ \text{mm}$. }
Os $L_{2,3}$-edge XAS and XMCD measurements were performed at the hard X-ray beamline BL39XU of SPring-8 \cite{suzukiBL39}. The magnetic field $B$ up to 7 T was applied parallel to the X-ray beam using a superconducting magnet. 
\red{
The measurements were done in the grazing incident geometry ($\sim$10$^\circ$ incidence angle). 
We note that the polycrystalline grain sizes were much smaller than the beam size ($\sim 0.4\ \text{mm} \times 0.4\ \text{mm}$). 
}
The circular polarization of the X rays was switched at each photon energy at a repetition rate of $\sim 30\ \text{Hz}$ using a diamond phase shifter. 
\red{The degree of the circular polarization of the X rays was better than 90\%. }
In order to eliminate spurious XMCD signals originating from differences in the optical paths of the two polarizations, the XAS and XMCD spectra taken at $B=\pm 7\ \text{T}$ were averaged. The sample was cooled using a closed-cycle refrigerator. The absorption signals were collected in the partial fluorescence-yield (PFY) mode using a silicon drift detector (SDD) 
\red{located nearly along the sample-normal direction (perpendicular to the incident X-ray direction)}. 
The Os $L_\alpha$ ($5d \to 2p_{\frac{3}{2}}$) and $L_\beta$ ($5d \to 2p_{\frac{1}{2}}$) fluorescence intensities were monitored for the Os $L_3$ ($2p_{\frac{3}{2}} \to 5d$) and $L_2$ ($2p_{\frac{1}{2}} \to 5d$) edges, respectively. No surface treatments prior to the measurements were made because the probing depth of the PFY mode  was sufficiently large ($\sim$ a few $\mu\text{m}$). The dead time of the SDD was corrected based on the nonparalyzable model \cite{arnold_nonparal, suzuki_PRB2005}. 

The XAS spectra were normalized so that the edge jump heights at the $L_3$ and $L_2$ edges, which were deduced from the center line of the extended X-ray absorption fine structure (EXAFS) oscillation, were equal to the ratio of 2:1. 
To compare the experimental XAS spectra [Fig.~\ref{xasxmcd}(a) and (b)] with the calculated ones, one needs to subtract the white-line (step-like) background and the EXAFS oscillation in the post-edge region. This has been done by fitting the experimental XAS spectra by the sum of two Lorentz functions and their integrals (arctangent functions), and then extracting the Lorentz function part. The results are shown in Fig.~\ref{xasxmcd}.
We note that, the uncertainties of the dead-time correction, spectral normalization by the edge jump heights, and the background-removal procedure described above, may lead to a total systematic error of $\sim10\%$ in the magnetic moments deduced from the XMCD sum rules [Eqs.\ (\ref{orbsumeq}), (\ref{spinsumeq})].

The orbital magnetic moment $M_\text{orb}$ and the `effective' spin magnetic moment $M_\text{spin}^{\text{eff}}$ have been deduced from the XAS and XMCD spectra using XMCD sum rules~\cite{OrbSum, SpinSum}: 
\begin{equation}
M_{\text{orb}}=-\frac{4\int_{L_3+L_2}(\mu^+-\mu^-)\text{d}\nu}{3\int_{L_3+L_2}(\mu^++\mu^-)\text{d}\nu}(10-N_d),
\label{orbsumeq}
\end{equation}
\begin{align}
& M_{\text{spin}}^{\text{eff}} \equiv  M_{\text{spin}}+\frac{7}{2}M_\text{T} \nonumber \\
& = -\frac{2\int_{L_3}(\mu^+-\mu^-)\text{d}\nu-4\int_{L_2}(\mu^+-\mu^-)\text{d}\nu}{\int_{L_3+L_2}(\mu^++\mu^-)\text{d}\nu}(10-N_d).
\label{spinsumeq}
\end{align}
Here, $\mu^+$ ($\mu^-$) is the XAS intensity for the positive (negative) helicity as a function of photon energy $h\nu$, 
$L_3$ ($L_2$) is the $2p_{3/2} \to 5d$ ($2p_{1/2} \to 5d$) absorption edge, 
and $N_d$ is the number of electrons in the $5d$ band.  \af{In the present work, $N_d$ is assumed to be the formal occupation number of 2 of the Os$^{6+}$ ($5d^2$) ion. Although $N_d$ is deviated from the formal occupation number due to hybridization with ligand (O $2p$) orbitals, we have ignored the deviation here because the absolute values of $M_{\rm orb}$ and $M_{\rm spin}$ are not important for the present purposes.} 
As shown in Eq.\ (\ref{spinsumeq}), the magnetic moment deduced from the spin sum rule 
is the sum of the spin magnetic moment $M_\text{spin}$ and an additional term $(7/2)M_{\text{T}}$ called `magnetic dipole'
defined as 
$\bm{M}_\text{T} \equiv -2 \sum_i \langle \bm{s}_i - 3(\bm{s}_i\cdot\bm{r}_i) \bm{r}_i/r_i^2 \rangle$, 
where $\bm{r}_i$ and $\bm{s}_i$ are, respectively, the position and the spin angular momentum operators of the $i$-th electron \cite{SpinSum,TXMCD_Stohr,TXMCD_Durr,shibata}. 
The magnetic dipole $\bm{M}_\text{T}$ represents the anisotropy of spin distribution and can be large in systems with a low symmetry or with strong SOC \cite{SpinSum,TXMCD_Stohr,TXMCD_Durr,shibata}. 
Thus, the difference between $M_{\text{spin}}^{\text{eff}}$ and $M_{\text{spin}}$ gives the degree of anisotropic spin-density distribution induced by the spin-orbit-entangled electronic structure. 

\section{Ligand-field multiplet calculation}\label{multiplet}

Ligand-field multiplet calculations were performed by using the XTLS 8.5 package \cite{tanakaxtls}. 
In general, the Slater integrals $F$'s and $G$'s (anisotropy of Coulomb interaction) and the SOC coupling constant $\zeta$ in solids are smaller than those of isolated atoms because the wavefunctions are spatially more extended due to hybridization. 
In order to model this effect, the atomic Slater integrals and $\zeta$, deduced from Hartree-Fock calculations \cite{Mann_SlaterInteg, Herman_SpinOrbitConst}, were multiplied by constant factors $R_\text{Slater}$ and $R_\text{SOC}$ ($0 \leq R_\text{Slater}< 1$, $0 \leq R_\text{SOC} < 1$), respectively. 
$R_\text{Slater}$ and $R_\text{SOC}$ and the cubic ligand-field splitting $\Delta_{\rm LF}$ and were treated as adjustable parameters. 
We used $\Delta_{\rm LF}= $ 4.3$\ \text{eV}$,  $\zeta = $ 0.33$\ \text{eV}$, and the reduction factor of 40 \% for the Slater integrals between the Os $5d$ orbitals. 
Note that the spin-orbit coupling for the $5d$ shell $\zeta$ and that for the $t_{2g}$ shell $\zeta'$ are related via $\zeta=-\zeta'$~\cite{kanamori}.
Hund's coupling $J_\text{H}$ between two $d$ electrons is related to 
Slater integrals through 
$J_\text{H} = \frac{3}{49}F^2 + \frac{20}{441}F^4=0.27$ eV~\cite{georges}. 
These parameter values are tabulated in Table~\ref{table1}.
We assumed that the incident X ray is parallel to the cubic [001] direction. We confirmed that the spectral line shape does not change in the case of cubic symmetry, as long as the incident X ray and the magnetic field are parallel. 

In the calculation of the Os $L_{2,3}$-edge XMCD spectra, the Zeeman energy due to the magnetic field $-\mu_\text{B}\bm{B}\cdot(\bm{L}+2\bm{S})$, where $\bm{L}$ and $\bm{S}$ are the orbital and spin angular momenta, respectively, was included.
The effect of the molecular field $-\mu_\text{B}\bm{H}_\text{mol}\cdot\bm{S}$, was incorporated by treating  $B=|\bm{B}|$ as an adjustable parameter and allowing the {\it B} value to exceed the external field of 7 T. 

\begin{table}
    \centering
    \caption{ 
    Parameter values for the Os $5d$ electrons hybridized with O $2p$ orbitals in Ba$_2$CaOsO$_6$ used in the present work.}
    \label{parameters}
    \begin{tabular}{lcc}
    \hline
      Parameter & Symbol & Value (eV) \\
    \hline
       ligand-field splitting & $\Delta_{\text{LF}}$ & 4.3\\
       Spin-orbit coupling for the $5d$ shell & $\zeta$ & 0.33 \\
       Hund's coupling & $J_{\rm H}$ & 0.27\\
   \hline
    \end{tabular}
    \label{table1}
\end{table}

The calculated spectra were broadened by a Lorentz function with the HWHM of 2.7 eV corresponding to the life time of the Os $L_{2,3}$ core hole ~\cite{Krause_corelifetime}. No Gaussian broadening was applied. 
As the initial state of the XAS and XMCD spectra, the five lowest states, i.e., the lowest $J_{\rm eff} = 2$ state in Fig.~\ref{ediagram}(b), were weighted according to the Boltzmann distribution and summed up.

\bibliography{ref}
\end{document}